Title: Aggregation of dipolar molecules in $SiO_2$ hybrid organic-inorganic films: use of silver nanoparticles as inhibitors of molecular aggregation

Article Type: Original research

Corresponding Author: Alfredo Franco, Ph.D.

Corresponding Author's Institution: Universidad Nacional Autónoma de México

First Author: Alfredo Franco, Ph.D.

Order of Authors: Alfredo Franco, Ph.D.;Jorge García-Macedo, Ph.D.;Giovanna Brusatin, Ph.D.;Massimo Guglielmi, Ph.D.

Abstract: Silver nanoparticles were synthesized with silver nitrate as precursor and Aminoethylaminopropyltrimethoxysilane as reducing agent. Their surface plasmon resonance was detected at 410 nm by UV-visible spectroscopy. The silver nanoparticles were immersed in $SiO_2$ hybrid sol-gel films doped with dipolar chromophores in order to study their effect as inhibitors of the dipolar chromophores aggregation. The presence of silver nanoparticles in the solid films was confirmed by Transmission Electronic Microscopy. The aggregation of the dipolar chromophores in the $SiO_2$ films was studied by UV-visible spectroscopy. Several organic molecules immersed in the films were tested too as inhibitors of aggregation. The effect of the silver nanoparticles on the aggregation of chromophores was compared with the effect due to the organic inhibitors. The results concerning to the inhibition of the chromophores aggregation are expressed in terms of the inhibitor, its concentration in the films and the temperature of the films during the UV-visible spectroscopy measurements.

# Aggregation of dipolar molecules in SiO$_2$ hybrid organic-inorganic films: use of silver nanoparticles as inhibitors of molecular aggregation

Alfredo Franco[1,2]*, Jorge García-Macedo[1], Giovanna Brusatin[2], Massimo Guglielmi[2]

[1]*Departamento de Estado Sólido, Instituto de Física, Universidad Nacional Autónoma de México, México, D.F. 04510 México.*

[2]*Dipartimento di Ingegneria Meccanica (Settore Materiali). Università degli Studi di Padova. Via Marzolo 9, 35131, Padova, Italia.*

*Corresponding author. Phone: +52(55)56225019

Fax: +52(55)56225011

e-mail: alfredofranco@fisica.unam.mx

Silver nanoparticles were synthesized with silver nitrate as precursor and Aminoethylaminopropyltrimethoxysilane as reducing agent. Their surface plasmon resonance was detected at 410 nm by UV-visible spectroscopy. The silver nanoparticles were immersed in SiO$_2$ hybrid sol-gel films doped with dipolar chromophores in order to study their effect as inhibitors of the dipolar chromophores aggregation. The presence of silver nanoparticles in the solid films was confirmed by Transmission Electronic Microscopy. The aggregation of the dipolar chromophores in the SiO$_2$ films was studied by UV-visible spectroscopy. Several organic molecules immersed in the films were tested too as inhibitors of aggregation. The effect of the silver nanoparticles on the aggregation of chromophores was compared with the effect due to the organic inhibitors. The results concerning to the inhibition of the chromophores aggregation are expressed in terms of the inhibitor, its concentration in the films and the temperature of the films during the UV-visible spectroscopy measurements.

AgNO$_3$, silver nitrate;

AEAPTMS, Aminoethylaminopropyltrimethoxysilane;

TEM, Transmission Electronic Microscopy;

DR1, Disperse Red 1;

ENPMA, Ethyl-[4-(4-nitro-phenylazo)-phenyl]-(2-oxiranylmethoxy-ethyl)-amine;

TEOS, Tetraethoxysilane;

GPTMS, (2-Glycidyloxypropyl)trimethoxysilane;

APTMS, [3-(2-Aminoethylamino)Propyl]trimethoxysilane;

PhTES, Triethoxyphenylsilane;

CbOH, 9H-Carbazole-9-ethanol;

Ph, Phenyl groups;

TS, Trans-Stilbene;

BPh, 4,4'-Bis(triethoxysilyl)-1,1'-biphenyl;

MeOH, Methanol;

MeOEtOH, Methoxyethanol.

# Introduction

The aggregation of chromophores in hybrid organic-inorganic films devoted to optics and photonics applications is a common problem to overcome (Lebeau and Innocenzi 2011). Chromophores with large dipolar moments, like the well known *N*-Ethyl-*N*-(2-hydroxyethyl)-4-(4-nitrophenylazo)aniline (commonly named Disperse Red 1 or simply DR1), are often embedded in sol-gel films for second order nonlinear optical applications, such as second harmonic generation and electro-optic modulation (Della Giustina et. al. 2006; Zhang et. al. 2006). These applications are possible only if the chromophores are arranged in a non-centrosymmetric way inside the films (Singer et. al. 1987).

In films used for second order nonlinear optical applications, it is a common practice to align the dipolar chromophores in a non-centrosymmetrical way through the application of an external strong electrostatic field (Giacometti et. al. 1999; Mortazavi et. al. 1989). These films are important in optical communications as part of electro-optical rectifiers (Dalton 2003). However there are two main problems to overcome in these films: to increase the amount of chromophores embedded in the materials and to increase the time the chromophores maintain their non-centrosymmetric arrangement; both challenges are intimately related to the chromophores large permanent dipole: the strong electrostatic dipole-dipole interactions induce the aggregation of the chromophores, breaking their non-centrosymmetric organization and diminishing the optical quality of the materials (Rau et. al. 2007). As the concentration of chromophores in the films increases, the dipole-dipole interactions become more evident (Reyes-Esqueda et. al. 2001), but it is necessary to have large amounts of chromophores in the films in order to attain large electro-optic $r_{33}$ and nonlinear second order optical coefficients $d_{33}$ (Singer et. al. 1987).

Several strategies have been followed in order to diminish these problems: to graft the dipolar chromophores to the matrix of the films (Li et. al. 2008; Matsuo et. al. 2009), to incorporate the chromophores into dendrimers (Dalton et. al. 2010) and to incorporate some appropriate organic molecules into the films (Chen et. al. 2008) help to reduce the spontaneous aggregation of the chromophores.

Carbazole molecules are among the organic molecules most commonly used as organic inhibitors of aggregation in the films. Carbazole moieties are particularly useful in photorefractive polymeric films, because they also work as electrical charge transporter (Grazulevicius et. al. 2003), but their role as inhibitors of the dipole-dipole interactions between chromophores in several kinds of films stands out in the works of Brusatin et. al. (2004), Marino et. al. (2006) and Reyes-Esqueda et. al. (2001).

We propose the incorporation of noble metal nanoparticles in the films as an alternative strategy to diminish the spontaneous aggregation of the dipolar organic chromophores. The high polarizability of the noble metal nanoparticles (Dupree and Smithard 1972; Strässler et. al. 1972) made them suitable for alternative ways to screen the dipolar interactions between chromophores (Franco et. al. 2011).

The research on noble metal nanoparticles has been intense and wide in the last decade due to their unique optical, electronic and chemical properties (Kamat 2002; Kelly et. al. 2003; Sardar et. al. 2009). Those properties made them ideal for the development of several kinds of optoelectronic, catalytic, or biosensing nanodevices (Kamat 2002; Moores and Goettmann 2006; Walters and Parkin 2009); many of them require the simultaneous presence of noble metal nanoparticles and organic molecules. These hybrid systems have been thoroughly studied through enhanced luminescence and Raman spectroscopies (Kamat 2002; Walters and Parkin 2009). But as much as we know there are not reports about the inhibiting effect of the noble metal nanoparticles on the aggregation of dipolar organic molecules. Specifically, we are interested in to study the effect of noble metal nanoparticles as inhibitors of dipole-dipole electrostatic interactions between dipolar chromophores, because some optical materials would improve their properties with their chromophores in a non-aggregated state; such kinds of materials can be, for example, films doped with luminescent dyes (Delbosc et. al. 2010; Halterman et. al. 2008), films doped with photoisomerizable molecules (Priimagi et. al. 2010) and films doped with second order non-linear optical molecules (Pizzotti et. al. 2009; Rau and Kajzar 2008).

In this work we have immersed silver nanoparticles in $SiO_2$ sol-gel films in order to study their effect on the aggregation of chromophores with large permanent dipolar moments. We report the aggregation inhibiting effect of several kinds of organic molecules as well as the aggregation inhibiting effect due to silver nanoparticles in a typical second order nonlinear optical system: $SiO_2$ hybrid organic-inorganic sol-gel films doped with DR1 chromophores. DR1 molecules tend to form H and J aggregates due to their large permanent dipolar moment (8.7 D) (Katz et. al. 1987), their aggregates are easily detected by UV-visible spectroscopy (Qiu et. al. 2010).

# Experimental

### Synthesis

Hybrid organic-inorganic $SiO_2$ films doped with dipolar push-pull chromophores and inhibitors were synthesized in order to study how noble metal nanoparticles influence the dipolar chromophores aggregation. The inhibitors were of two types: organic inhibitors or silver nanoparticles. The studied dipolar chromophore was the well known push-pull molecule 2-{Ethyl-[4-(4-nitro-phenylazo)-phenyl]-amino}-ethanol, commonly named Disperse Red 1 (DR1). The large permanent dipolar moment of these molecules (8.7 D) favors their molecular aggregation. The doped $SiO_2$ films were synthesized by the sol-gel method and deposited on soda lime substrates by spin-coating at 1000 rpm for 30 seconds.

Some films were doped with a modified DR1 molecule too. The modification done to the DR1 molecules is published elsewhere (Franco et. al. 2010), these molecules are the Ethyl-[4-(4-nitro-phenylazo)-phenyl]-(2-oxiranylmethoxy-ethyl)-amine (ENPMA). The ENPMA molecules were used because, in contrast to DR1, they can be easily grafted to the $SiO_2$ backbone of the hybrid films.

Each one of the studied films were synthesized under slightly different procedures, depending on the matrices, chromophores and molecular inhibitors in the films.

Several alcoxides were used as sol-gel precursors: Tetraethoxysilane (TEOS), (2-Glycidyloxypropyl)trimethoxysilane (GPTMS), [3-(2-Aminoethylamino)Propyl]trimethoxysilane (AEAPTMS), and Triethoxyphenylsilane (PhTES).

Several moieties were tested as organic molecular inhibitors: 9H-Carbazole-9-ethanol (CbOH), Phenyl groups (Ph), trans-1,2-Diphenylethylene (also known as Trans-Stilbene or simply TS) and 4,4'-Bis(triethoxysilyl)-1,1'-biphenyl (BPh).

All the reactants used for the synthesis were purchased from Aldrich.

### Synthesis of films without inhibitors

The synthesis of films without inhibitors was carried out in several steps. First, GPTMS and TEOS were prehydrolized by mixing GPTMS, TEOS, water and methanol (MeOH) in a [TEOS:GPTMS:H$_2$O:MeOH] molar ratio equal to [1:2.33:6.66:18.33].

GPTMS and TEOS were stirred together for 10 minutes at room temperature, then water was added dropwise, followed by methanol.

This solution was stirred magnetically for 10 minutes. Finally, it was refluxed during 4 hours at 80°C. The final sol was named GT; its theoretical SiO$_2$ concentration is 2.08 moles/L.

On the other hand, the chromophores (DR1 or ENPMA) were dissolved in Methoxyethanol (MeOEtOH), then AEAPTMS was added, and the solution was stirred for 5 minutes. This solution was then mixed with the GT sol, then it was stirred at 40°C during one night. Finally, the film deposition was done.

For the sake of clarity, the missing molar ratios in this synthesis are specified some lines below.

### Synthesis of films with organic inhibitors.

Two different procedures, based in the previous one, were used for the synthesis of films with organic inhibitors. One of them follows the same procedure described for the films without inhibitors but, just before the incorporation of AEAPTMS, it was added CbOH, TS or BPh to the solution. This solution was then stirred for 15 minutes, thus the AEAPTMS was added and the procedure followed as the one already described above.

The other procedure consisted in to use PhTES instead of TEOS, in this case the final sol was named GPhT instead of GT. The main difference between both sols is the presence of phenyl groups attached to the SiO$_2$ in the GPhT sol. In this way the phenyl groups can be tested as molecular inhibitors too. The rest of the synthesis was analogue to that one for the films without molecular inhibitors. For the sake of clarity, the specific molar ratios used for this synthesis are specified some lines below.

### Synthesis of films with silver nanoparticles.

In the case of the films with silver nanoparticles, the procedure consisted in to add 5 μl of AEAPTMS to 0.13 mg of silver nitrate (AgNO$_3$) previously dissolved in 800 μl of MeOEtOH in a flask. The solution was stirred ultrasonically during 15 minutes. Then DR1 was added and everything was stirred magnetically at room temperature for 15 minutes. At last, 30.3 μl of GT were added. The final solution was stirred magnetically during 30 minutes at 40°C and the film deposition was done.

Along the different synthesis, the AEAPTMS plays two different roles: (a) it forms covalent bonds through the epoxy rings of the GPTMS and ENPMA molecules and (b) it behaves as a reducing agent, converting silver ions into silver colloids.

### Studied films

For a clear identification of the films, they were labeled in accordance to their constituents and their molar ratios. The notation of their labels is as follows: *M/C.X/S(or T).Y*, with *M* the prehydrolized sol (GT or GPhT); *C* the hosted chromophore (DR1 or ENPMA); *X* the chromophores molar percentage with respect to $SiO_2$ (05, 10, 15, 20, 30 or 40%); *S* the organic molecular inhibitor (CbOH, TS or BPh), if any; and *Y* the molar ratio between the molecular inhibitors and the chromophores (00, 01, 02, 03 or 04). On the other hand, *T* indicates the temperature of the film at which the UV-visible spectra were recorded (20, 70, 90, or 110°C).

Thus, the molar ratio between the components of the set of samples labeled as GT(GPhT)/DR1(ENPMA).*X* is:
[DR1(ENPMA):AEAPTMS:GT(GPhT):MeOEtOH]=[1:1:x:373], with x equal to 19.00, 9.00, 5.67, 4.00, 2.33 or 1.50 for *X* equal to 05, 10, 15, 20, 30 or 40, respectively.

In the same way, the set of samples labeled as GT/DR1(ENPMA).*X*/CbOH(TS or BPh).*Y* have the next molar ratios between their components:
[DR1(ENPMA):AEAPTMS:GT:CbOH(TS or BPh):MeOEtOH]=[1:1:x:y:373], with x equal to 19.00 or 2.33 for *X* equal to 05 or 30, respectively; and y equal to 0, 1, 2, 3 or 4 for *Y* equal to 00, 01, 02, 03 or 04, respectively.

The DR1 molecules inside the films doped with silver nanoparticles were in a molar percentage, with respect to $SiO_2$, equal to 15, 30 and 45%.

Fig. 1 shows all the reactants used during the synthesis of the samples.

Fig. 2 shows a scheme with the general procedure followed for the synthesis of the films.

## Measurements

The influence of the inhibitors on the chromophores aggregation as well as the formation of the silver nanoparticles were studied by UV-visible optical absorption spectroscopy, for several kinds of inhibitors and for several concentrations of the inhibitors in the films.

We do not expect a strong aggregation of the chromophores when the concentration and the polarizability of the inhibitors are high enough.

The UV-visible measurements were performed on each film, just after their deposition, using a JASCO V-570 spectrophotometer in transmission mode. The spectrophotometer sample holder allows to control the temperature of the film at which the spectrum is recorded.

The films with silver nanoparticles were also studied by transmission electronic microscopy (TEM) with a JEM-2010F FASTEM microscope in order to confirm the existence of metal nanoparticles in the films.

## Results and discussion

The results show the aggregation states of the films as function of the relative concentrations of the inhibitors with respect to the dipolar molecules. It is assumed hereafter that UV-visible spectroscopy is able to reflect the state of aggregation of the chromophores, as reported for chromophores with a large dipolar moment embedded in films (Brusatin et. al. 2004; Choi et. al. 2000; Marino et. al. 2008; Priimagi et. al. 2005).

In general, the UV-visible spectra of the samples consist of a main band due to the electronic transitions in the double conjugated bonds system of the chromophores. The main band is superimposed to a baseline due to the substrate and to the inorganic matrix of the films, whose absorbance increases at shorter wavelengths. The differences between the several UV-visible spectra of these samples are mainly due to (a) the appearance of a secondary peak at 404 nm which reveals that the chromophores are aggregated (Priimagi et. al. 2005) and to (b) the peaks at 332 nm and 345 nm which evidences the presence of CbOH moieties in some of the samples (Brusatin et. al. 2004).

The spectra of the GT/DR1.X samples without silver nanoparticles are shown in Fig. 3 (a). These spectra show, as expected, that an increment in the load of the chromophores favors the formation of aggregates in the films. The aggregation of the chromophores is larger as their concentration increase; in such case the intensity of the peak at 404 nm increases too. It is useful to consider the empirical parameter $\delta$ in order to quantify the aggregation of the chromophores. This empirical parameter considers the height of the peak at 404 nm as evidence of the degree of the chromophores aggregation; as suggested by Ghosh et. al. (2008), it is given by:

$$\delta = 1 - \frac{(O.D._{main} - O.D._{404})}{[O.D._{main} - O.D._{404}]_0}, \quad (1)$$

where $[O.D._{main}-O.D._{404}]_0$ is the difference between the optical densities at the main absorption band and the optical density at the 404 nm peak in a blank sample with its shortest possible value at 404 nm; similarly, $(O.D._{main}-O.D._{404})$ is the same difference for the sample, but for any other height at 404 nm. Thus, $\delta$ is zero for the sample with the smallest peak at 404 nm, and $\delta$ increases when aggregation also increases.

In Fig. 3 (b) $\delta$ is plotted as function of the chromophores concentration, this plot evidences that larger aggregation occurs as the density of chromophores in the films increases.

The chromophores aggregation is related to the temperature of the samples; thermal energy provides to the chromophores enough mobility to break the aggregates (Kim et. al. 2001). The UV-visible spectra of the GT/DR1.05/T.X samples were taken at several $X$ temperatures; they are shown in Fig. 4. As can be seen from Fig. 4, an increase in the temperature of the sample induces a decrease in the height of the peak at 404 nm. It confirms that the 404 nm peak is effectively associated to the aggregation of the dipolar molecules.

The same effect observed in the GT/DR1.05/T.X films was detected in films with different chromophores concentrations (Fig. 5), but the temperatures needed to break the aggregates in the films are higher for larger chromophores concentrations.

The covalent linkage of the chromophores to the matrix of the films helps to reduce the aggregation, because the mobility of the dipolar molecules is reduced, thus the linkage avoids the free electrostatic coupling of the chromophores. Fig. 6 shows the spectra of the samples with ENPMA molecules. The ENPMA chromophores are linked to the matrix through their opened epoxy ring. From Fig. 6 it can be seen that aggregation is indeed reduced, but it is not completely avoided. In fact, there is not a clear aggregation up to a 20% of the chromophores concentration in the films; however, in the sample with a 30% of chromophores concentration the band at 404 nm is, instead, clearly visible. In Fig. 6, the spectrum of a sample simultaneously doped with 30% of chromophores and CbOH moieties exhibits how efficient this organic inhibitor is; the peak at 404 nm disappears because the CbOH molecules avoid the aggregation.

There is another difference between systems with and without grafted chromophores. In the case of the DR1 molecules, the synthesis procedure does not affect their aggregation, which is, instead, avoided by the inhibitors. But at low concentrations ENPMA molecules, it is not possible to distinguish between the effect of the inhibitor and the effect of the linkage to the matrix. It means that at high chromophores concentrations it is not enough to link the chromophores to the matrix of the films to avoid their aggregation, and the use of inhibitors is still necessary.

In order to study the effect of inhibitors in the aggregation of the chromophores, the work was focused on DR1 guest-host systems, because in these systems the peak at 404 nm is clearly visible even at low concentrations.

In general, samples with low chromophores concentrations are less affected by aggregation due to the large distance between the chromophores inside the homogeneous films, the electrostatic interactions between permanent dipoles become almost negligible, which is straightforward understandable from the fact that dipole-dipole forces between two molecules vary inversely as the third power of the distance between the molecules (Israelachvili 2002). Thus the UV-visible spectra of the samples with the lowest concentration of chromophores (GT/DR1.05/T.110, GPhT/DR1.05 and GT/ENPMA.05) are good references of negligible aggregation systems. Even if small changes in the matrix composition may affect the shape of the main chromophore band, this effect is small compared to the effect due to aggregation.

Fig. 7 shows the spectra of the DR1 guest-host films. It is possible to detect the dependence of the chromophores aggregation on the type and concentration of the inhibitor molecules from the spectra in Fig. 7. Three different kinds of molecules were added to the films as organic inhibitors: BPh, TS and CbOH. Several concentrations of the inhibitors were tested. The matrix precursor PhTES behaves as an organic inhibitor too, due to its polarizable phenyl group. As the amount of PhTES in the films was fixed, the effect of the four organic molecules on the aggregation of the chromophores was evaluated by considering the chromophore/inhibitor ratio in each case. Therefore, in Fig. 7 (a) the spectra of the samples containing PhTES and different amounts of DR1 are shown; the PhTES/DR1 ratio ranges between 0.45 (GphT/DR1.40 sample) and 5.70 (GphT/DR1.05 sample). In Fig. 7 (b) the spectra of the samples containing 5% of DR1 and BPh as inhibitor are shown; the BPh/DR1 ratio ranges between 0 (GT/DR1.05/BPh.00 sample) to 4 (GT/DR1.05/BPh.04 sample). Fig. 7 (c) shows the spectra of samples containing 5% of DR1 and CbOH as inhibitor; the CbOH/DR1 ratio ranges between 0 (GT/DR1.05/CbOH.00 sample) to 4 (GT/DR1.05/CbOH.04 sample). In Fig. 7 (d) the spectra of samples containing 5% of DR1 and TS as inhibitor are shown; the TS/DR1 ratio ranges between 0 (GT/DR1.05/TS.00 sample) to 4 (GT/DR1.05/TS.04 sample). Finally, Fig. 7 (e) shows the spectra of samples with a higher content of DR1 chromophores (30%) and CbOH as inhibitor; the CbOH/DR1 ratio ranges between 0 (GT/DR1.30/CbOH.00 sample) to 4 (GT/DR1.30/CbOH.04 sample).

The $\delta$ parameters corresponding to the spectra in Fig. 7 are shown in Fig. 8 as function of the inhibitor/DR1 molar ratio.

From Fig. 8 it is clear that the aggregation decreases as the inhibitor concentration increases; actually, for the same chromophores concentration it seems that the inhibitors are more effective as their polarizability increases, as Table 1 suggests. It means that the amounts $N$ of inhibitors needed to avoid the aggregation of the dipolar chromophores increases following the next order: $N$(CbOH) < $N$(TS) < $N$(BPh) < $N$(Ph). In case that larger concentrations of chromophores be present in films, larger inhibitor/chromophore ratios will be needed, as confirmed by the results obtained for the GT/DR1.30/CbOH.$X$ samples. These results suggest the use of inhibitors with large polarizabilities in the films; they should be able to avoid the aggregation of dipolar chromophores in a more effective way. We tested the performance of silver nanoparticles as inhibitors, because their large polarizabilities (Dupree and Smithard 1972; Strässler et. al. 1972).
Silver nanoparticles were synthesized as described in the experimental section above. In solution, they shows the UV-visible spectrum of the Fig. 9. Its band is centered at 410 nm, due to the surface plasmon resonance of the silver nanoparticles, fingerprint of the existence of silver nanoparticles, as reported elsewhere (Moores and Goettmann 2006; Renteria and García-Macedo 2005). The films doped with silver nanoparticles were studied by TEM too, in order to confirm the presence of silver nanoparticles in the films. The micrographs presented in Fig. 10 correspond to nanoparticles in a GT/DR1.30 film. They show a broad distribution of quasi-spherical silver nanoparticles in the films, with lattice parameters of 2.03 Å and 2.05 Å corresponding to (2 0 0) silver planes.
The hybrid films doped simultaneously with a constant concentration of silver nanoparticles and several concentrations of DR1 chromophores exhibit the UV-visible spectra shown in Fig. 11(a). As observed from the spectra there is aggregation of the chromophores only at very high concentrations. It is expected to have less aggregation in these films than in the other ones due to the high polarizability of the silver nanoparticles. As reported in a previous work (Franco et. al. 2011), the high polarizability of the silver nanoparticles screens the electrostatic dipole-dipole interactions. The presence of the peak at 404 nm indicates the formation of aggregates, but the surface plasmon resonance peak at 410 nm is not visible because the main band, due to the chromophores, covers it. Fig. 11 (b) shows the $\delta$ parameter as function of the DR1/SiO2 ratio for a constant concentration of silver nanoparticles, instead as function of DR1/silver nanoparticles ratio, because it is not straightforward to quantify the amount of silver nanoparticles formed from the original concentration of silver nitrate. As can be seen form Fig. 11 (b), there is not almost aggregation at the 30% chromophores concentration; the films preserve their high optical quality. It makes to silver nanoparticles a new kind of efficient inhibitor of aggregation, and a good alternative to the commonly used CbOH organic inhibitor.

## Conclusions

The UV-visible spectra of $SiO_2$ hybrid films doped simultaneously with dipolar DR1 chromophores and polarizable organic molecules showed the effect of the polarizable molecules on the chromophores aggregation. The organic molecules tested as molecular inhibitors reduced the dipole-dipole interactions between chromophores with large dipolar moments. The dipole-dipole interactions between chromophores were better screened as the concentration of the inhibitors or their polarizabilities were larger. Silver nanoparticles also were tested as inhibitors of the dipolar chromophores aggregation, due to their high polarizability. Their performance as inhibitors of the aggregation was sufficiently good to consider them as an alternative to the use of organic polarizable molecules in films doped with high loads of dipolar chromophores.

## Acknowledgements

The authors thank to Diego Quiterio for the preparation of samples for TEM studies and to Roberto Hernández-Reyes for TEM technical assistance. The authors also thank to CONACYT 79781, NSF-CONACYT, PUNTA, RedNyN, PAPIIT IN107510, FIRB Italian project

RBNE033KMA"Molecular compounds and hybrid nanostructured material with resonant and non resonant optical properties for photonic devices" and UNAM-UNIPD agreement for financial support.

# Figure legends

Fig. 1 Molecules used for the synthesis of the films.

Fig. 2 Scheme of the procedure followed for the synthesis.

Fig. 3 (a) UV-visible absorbance spectrum of GT/DR1.X films. (b) $\delta$ parameter as function of the DR1 concentration.

Fig. 4 (a) UV-visible absorbance spectrum of GT/DR1.05/T.X films. (b) $\delta$ parameter as function of the temperature.

Fig. 5 (a) UV-visible absorbance spectrum of GT/DR1.X/T.Y films. The continuous plots correspond to 20ºC, the dashed plots correspond to 110ºC. (b) $\delta$ parameter as function of the DR1 concentration.

Fig. 6 (a) UV-visible absorbance spectrum of GT/ENPMA.X films and the GT/ENPMA.30/CbOH.03. (b) $\delta$ parameter as function of the ENPMA concentration.

Fig. 7 UV-visible absorbance spectrum of: (a) GPhT/DR1.*X*, (b) GT/DR1.05/BPh.*X*, (c) GT/DR1.05/CbOH.*X*, (d) GT/DR1.05/TS.*X*, (e) GT/DR1.30/CbOH.*X* films.

Fig. 8 $\delta$ parameter as function of the inhibitor:DR1 molar ratio.

Fig. 9 UV-visible absorbance spectrum of silver nanoparticles in MeOEtOH. The band at 410 nm is the fingerprint of the nanoparticles, it is due to their surface plasmon resonance.

Fig. 10 Representative TEM micrographs of the GT/DR1.*X* films with silver nanoparticles.

Fig. 11 (a) UV-visible absorbance spectrum of GT/DR1.X films with silver nanoparticles. (b) $\delta$ parameter as function of the DR1 concentration.

# Table

Table I. Estimated polarizability values for the organic molecules tested as inhibitors

(the values were calculated through ACDlabs freeware)

|  | PhTES | BPh | TS | CbOH |
|---|---|---|---|---|
| Molecule polarizability [x $10^{-24}$ cm$^3$] | 10.4 | 20.1 | 23.3 | 25.5 |

Figure 01
Click here to download line figure: figure01.eps

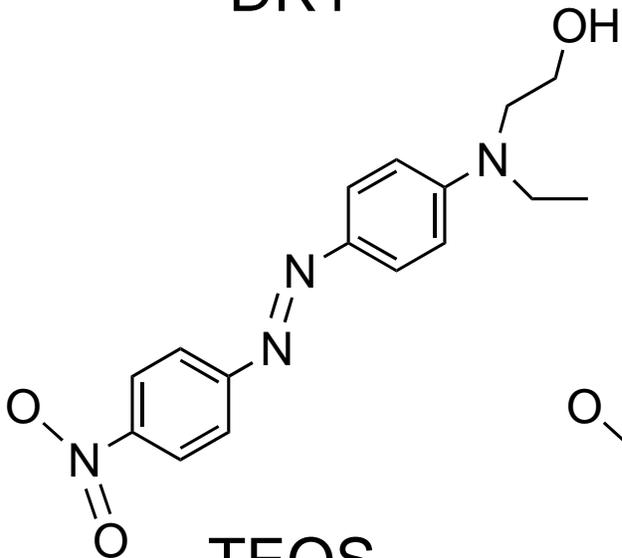
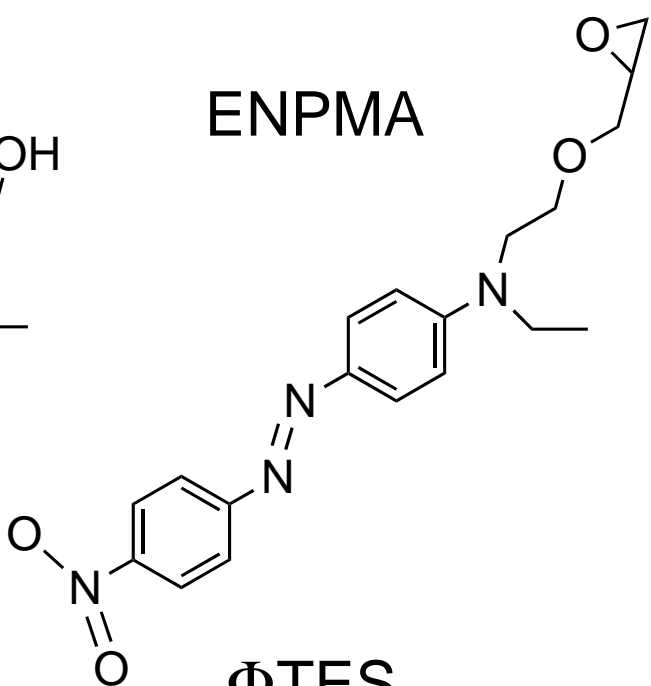
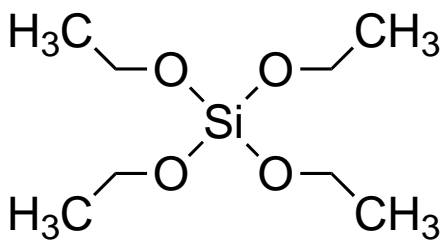
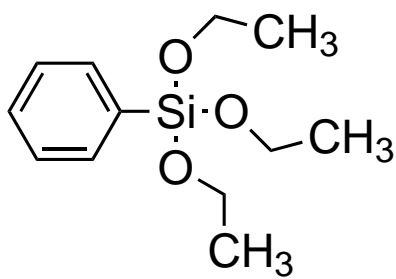
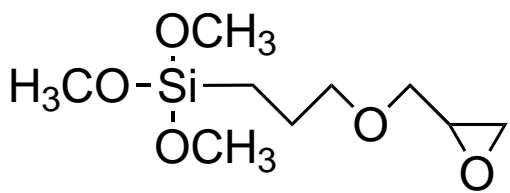
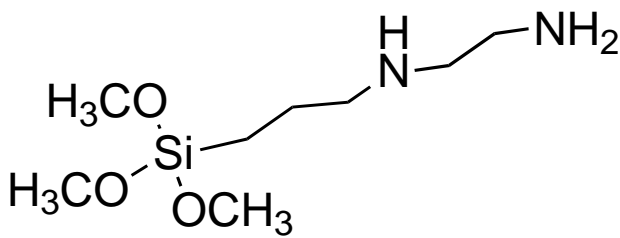
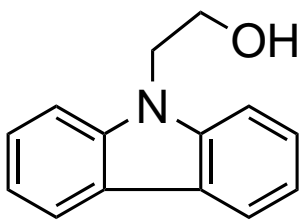
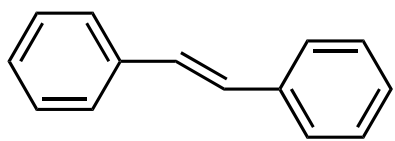
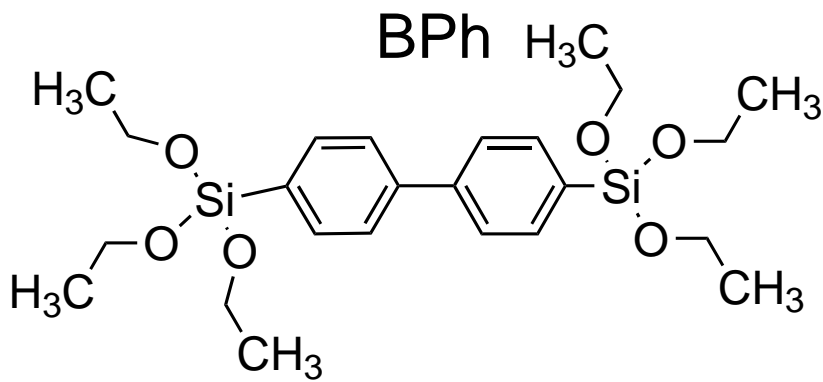

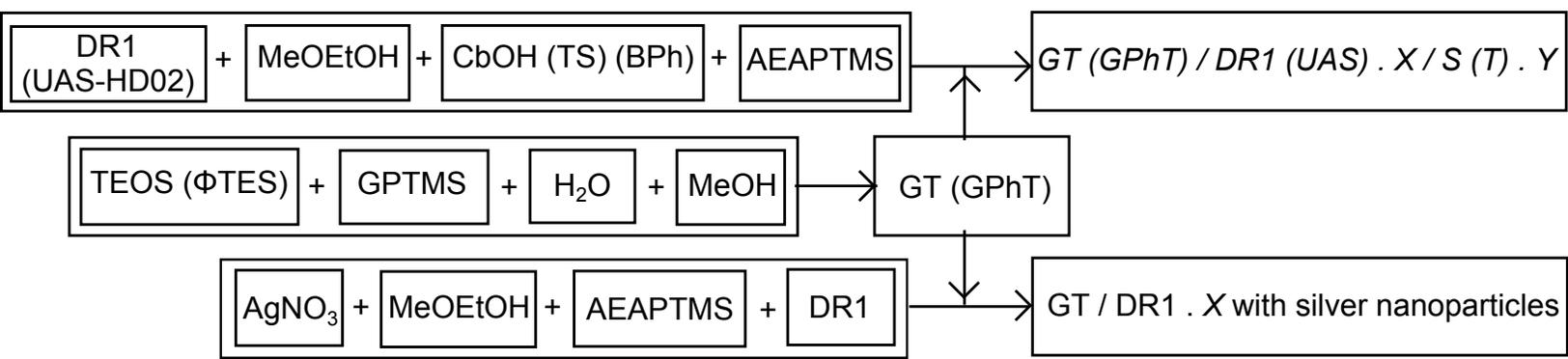

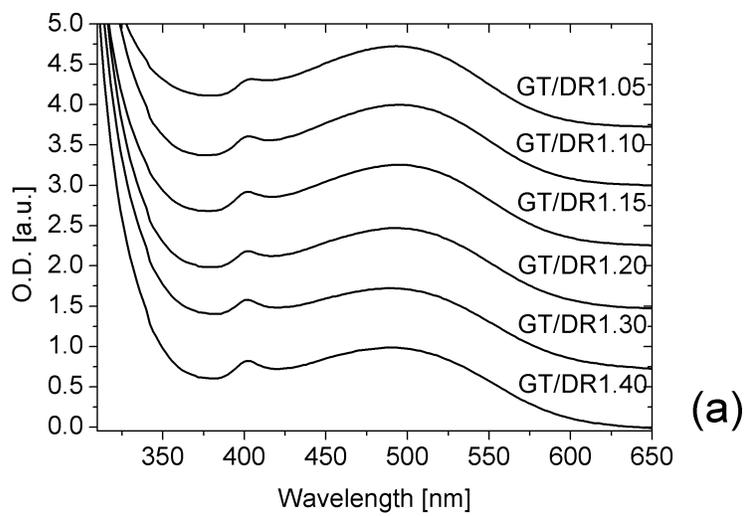 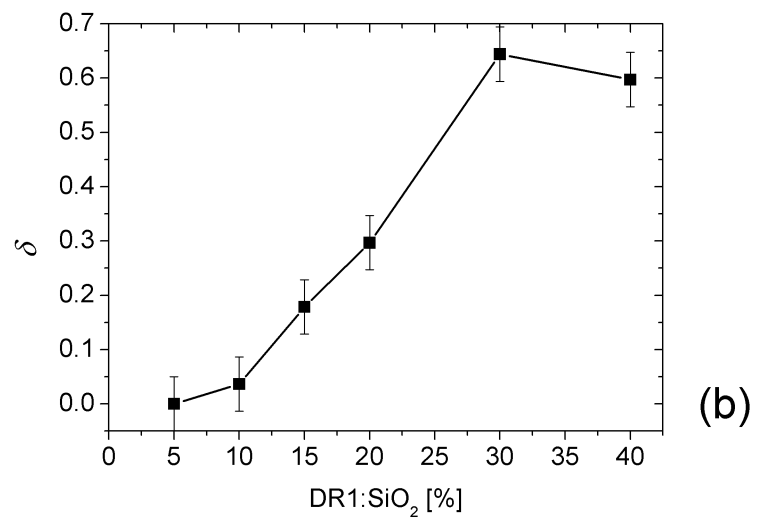

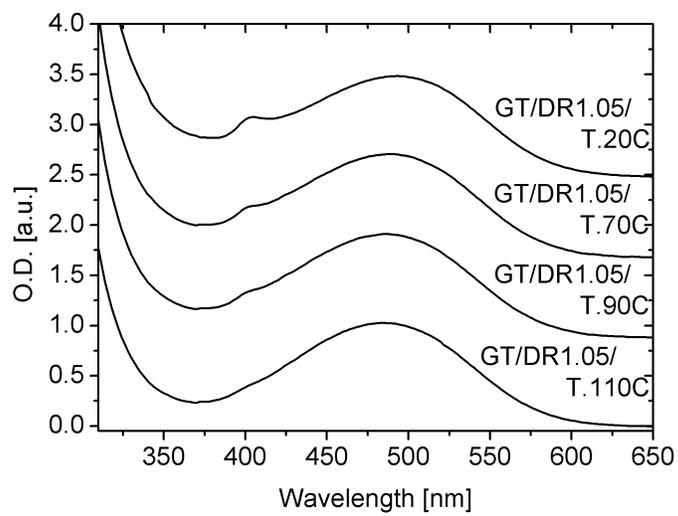
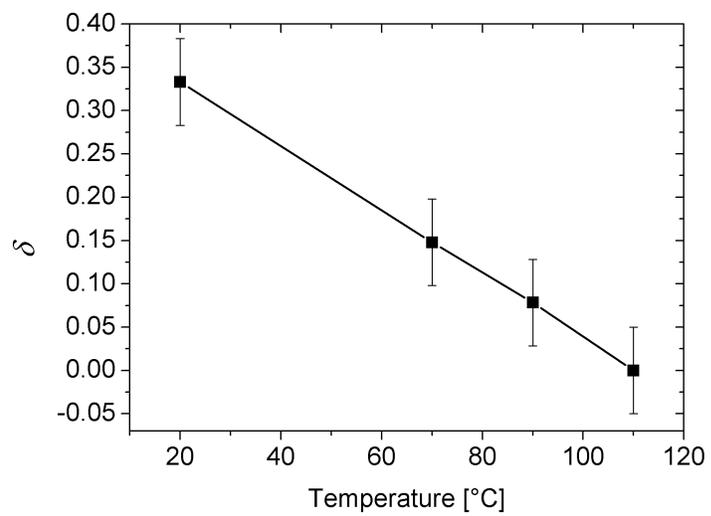

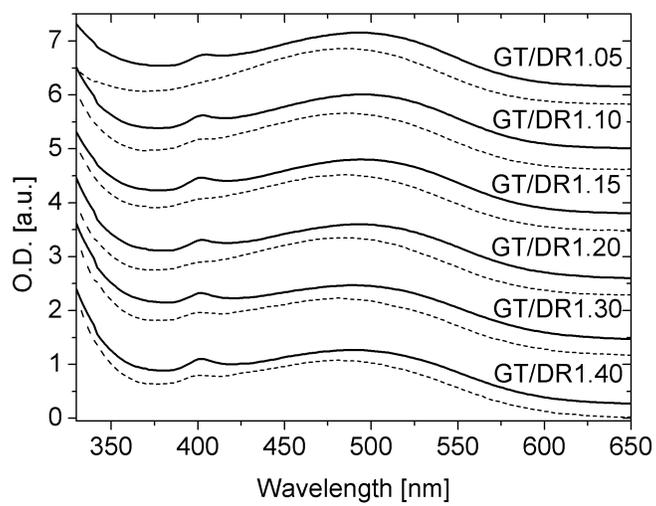 (a)
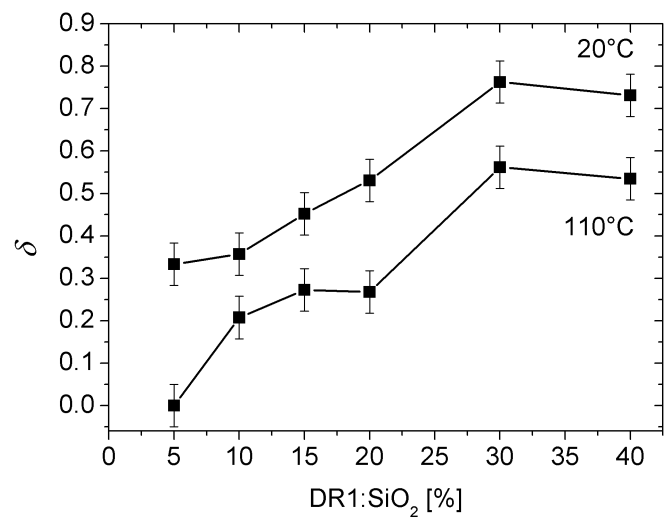 (b)

**Figure 06**
Click here to download line figure: figure06.eps

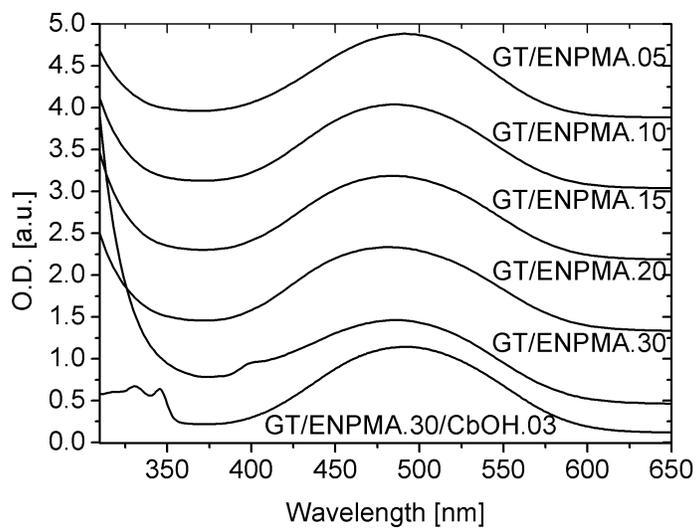 (a)

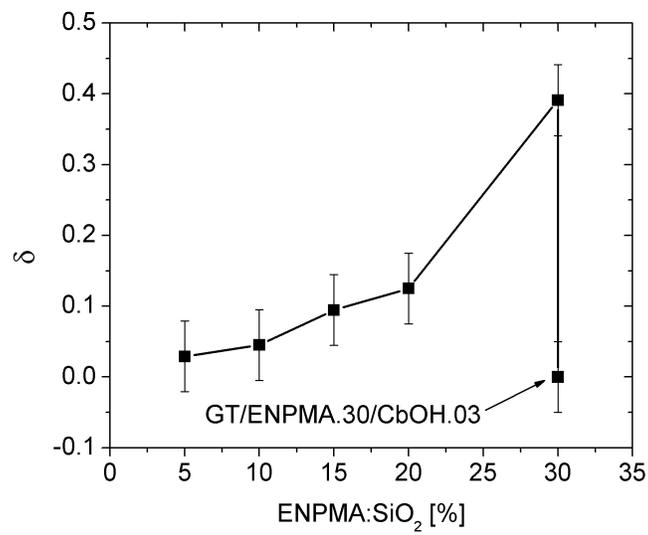 (b)

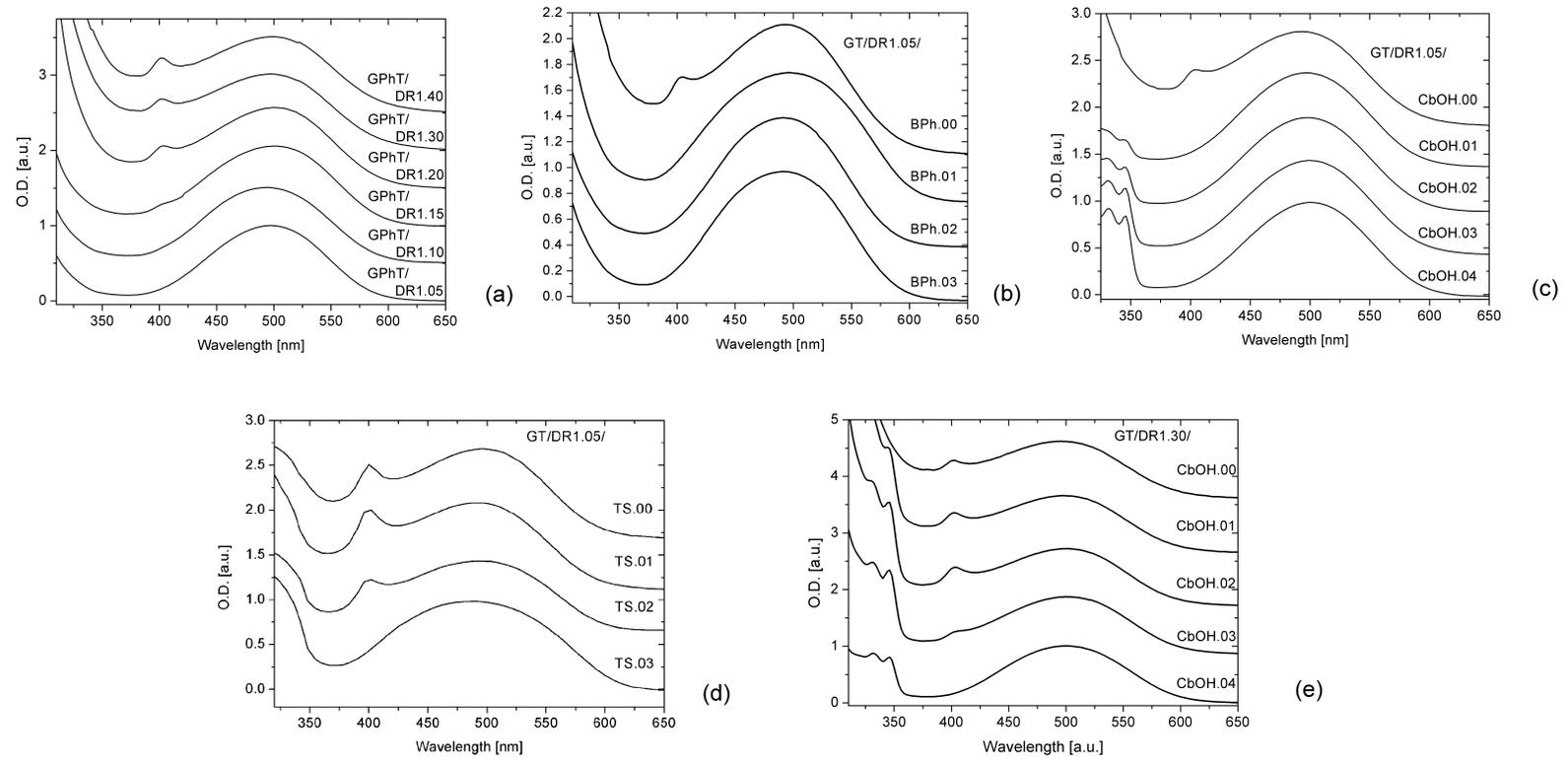

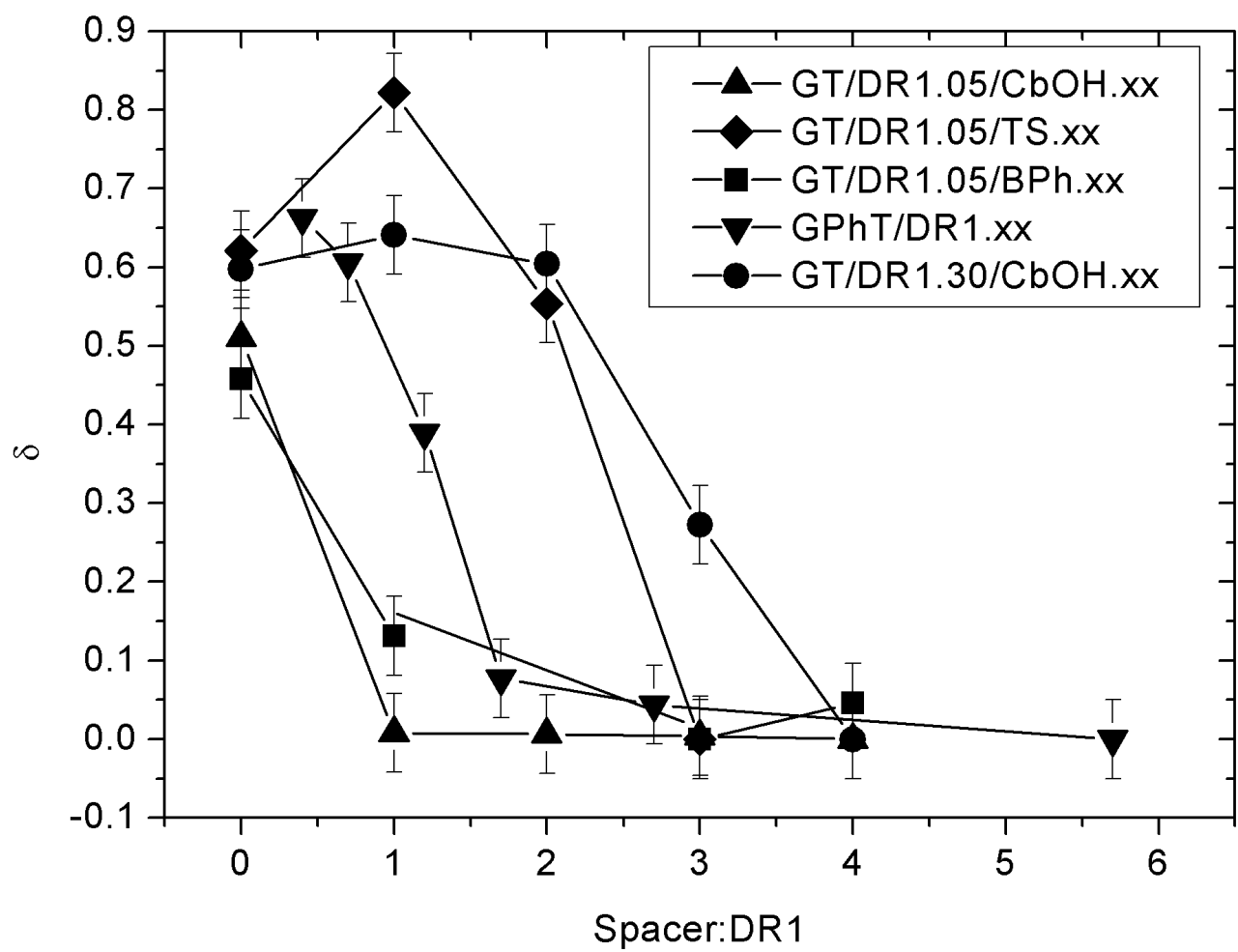

**Figure 09**
Click here to download line figure: figure09.eps

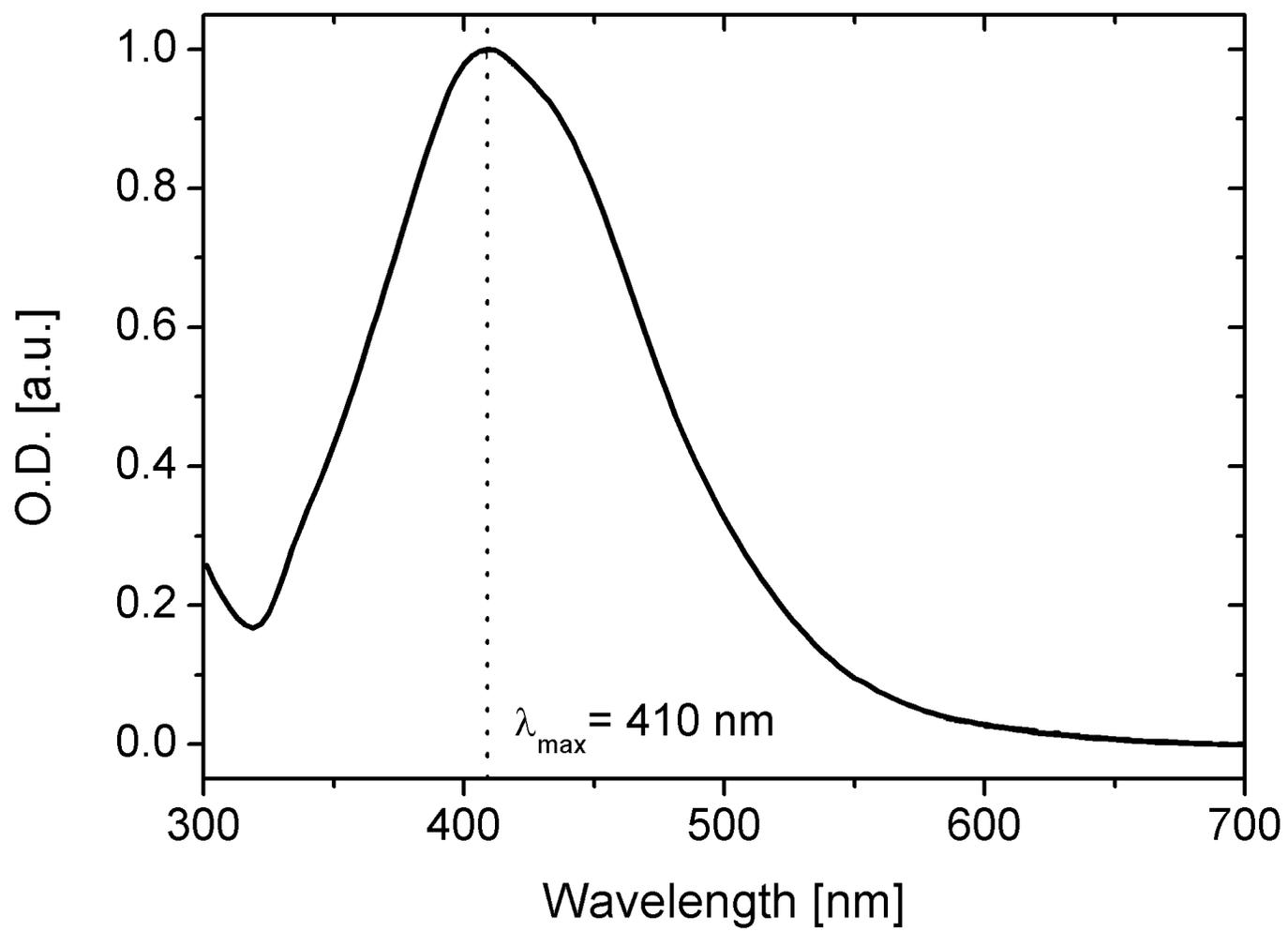

**Figure 10**
Click here to download high resolution image

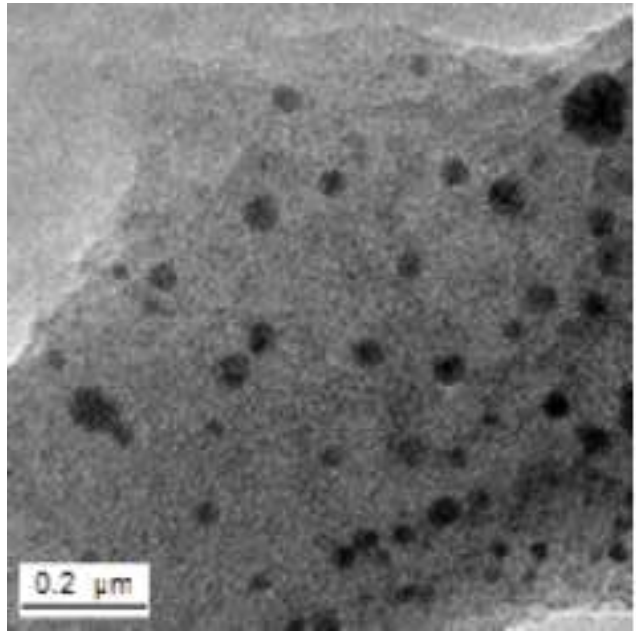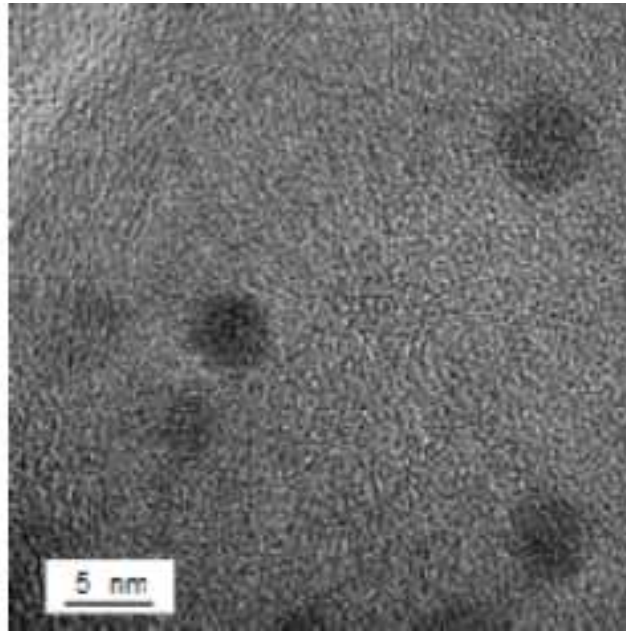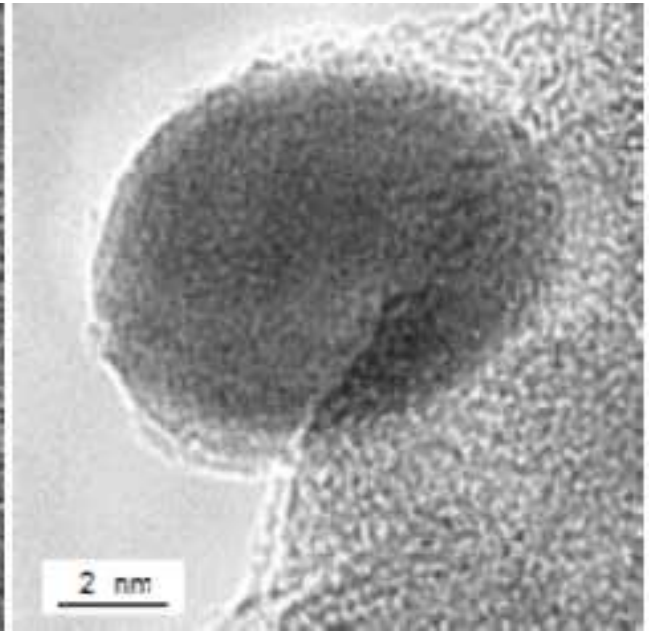

**Figure 11**
Click here to download line figure: figure11.eps

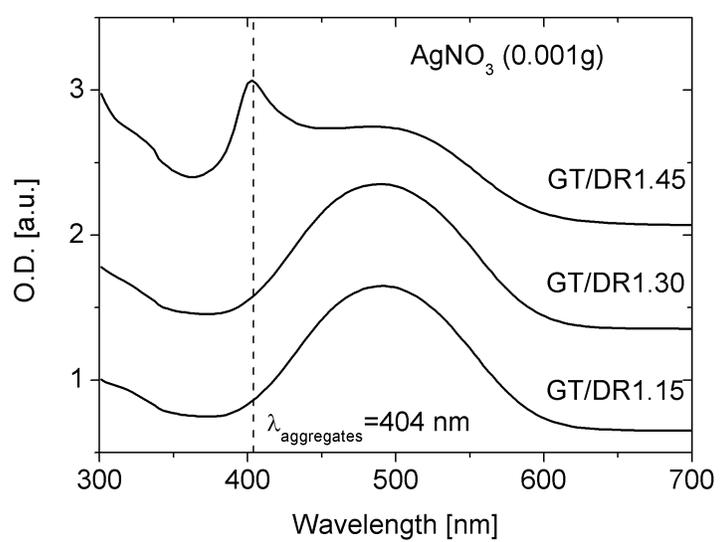 (a)

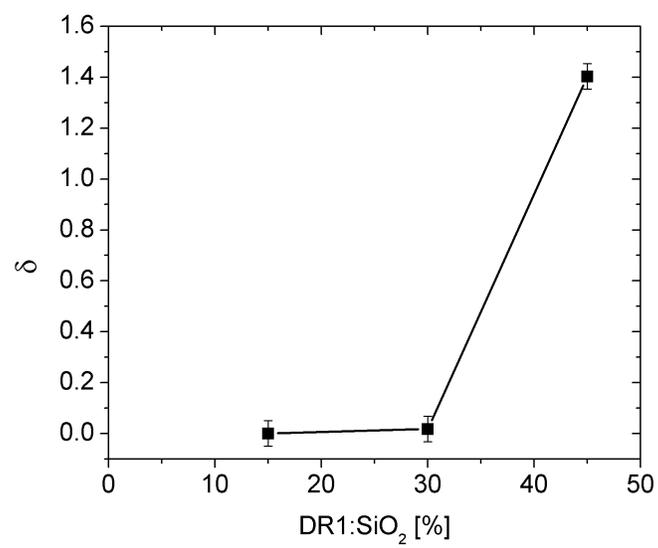 (b)